\documentclass{aastex631}

\usepackage{amsmath}
\usepackage[version=3]{mhchem} 

\begin{document}

\title{Moist convection is most vigorous at intermediate atmospheric humidity}

\author[0000-0003-0769-292X]{Jacob T. Seeley}
\affiliation{Department of Earth and Planetary Sciences \\
Harvard University \\
Cambridge, MA 02461, USA}

\author[0000-0003-1127-8334]{Robin D. Wordsworth}
\affiliation{Department of Earth and Planetary Sciences \\
Harvard University \\
Cambridge, MA 02461, USA}



\begin{abstract}

In Earth's current climate, moist convective updraft speeds increase with surface warming. This trend suggests that very vigorous convection might be the norm in extremely hot and humid atmospheres, such as those undergoing a runaway greenhouse transition. However, theoretical and numerical evidence suggests that convection is actually gentle in water vapor-dominated atmospheres, implying that convective vigor may peak at some intermediate humidity level. Here, we perform small-domain convection-resolving simulations of an Earth-like atmosphere over a wide range of surface temperatures and confirm that there is indeed a peak in convective vigor, which we show occurs near $T_s\simeq330$ K. We show that a similar peak in convective vigor exists when the relative abundance of water vapor is changed by varying the amount of background (non-condensing) gas at fixed $T_s$, which may have implications for Earth's climate and atmospheric chemistry during the Hadean and Archean. We also show that Titan-like thermodynamics (i.e., a thick nitrogen atmosphere with condensing methane and low gravity) produce a peak in convective vigor at $T_s\simeq95$ K, which is curiously close to the current surface temperature of Titan. Plotted as functions of the saturation specific humidity at cloud base, metrics of convective vigor from both Earth-like and Titan-like experiments all peak when cloud-base air contains roughly 10\% of the condensible gas by mass. Our results point to a potentially common phenomenon in terrestrial atmospheres: that moist convection is most vigorous when the condensible component is between dilute and non-dilute abundance.

\end{abstract}



\section{Introduction} \label{sec:intro}
Moist convection produces some of the most impactful weather and climate phenomena on Earth, from reflective stratocumulus layers to drenching seasonal monsoons. A longstanding challenge in climate science is to improve the crude treatment of such convective phenomena in general circulation models, which hampers predictions of critical quantities such as changes in local precipitation and the sensitivity of Earth's climate to CO$_2$. Looking beyond Earth, there is strong evidence that moist convection --- that is, convection coupled to phase changes of a condensible substance --- factors into the past and present evolution of the majority of solar system atmospheres, including Venus \citep{Kasting1988}, Mars \citep{Wordsworth2016,Yamashita2016}, Jupiter \citep{Gierasch2000}, Saturn \citep{Li2015}, the ice giants \citep{Hueso2020}, and Titan \citep{Schneider2012}. Therefore,  in  order  improve  our  understanding  of  contemporary Earth  climate  and  planetary climate more generally,  there  is  a  clear  need  to  deepen  and  generalize  our  understanding  of moist convective physics.

In this paper, we take a step toward a more generalized understanding of moist convection by focusing on one of its most basic characteristics: convective vigor, or the buoyancy and vertical velocity of cloud updrafts \citep[e.g.,][]{Zipser2006,Hansen2015,Hansen2020}. Updraft speeds affect instantaneous surface precipitation rates \citep{Muller2020} and the fraction of cloud condensate that is lofted instead of falling to the surface, which in turn exerts a strong influence on cloud cover and climate sensitivity \citep{Zhao2014,Zhao2016}. More vigorous updrafts are more likely to overshoot the tropopause and inject near-surface air into the stratosphere, thereby affecting stratospheric humidity and chemistry \citep{Liu2020,ONeill2021} and planetary water loss on geologic timescales \citep{Wordsworth2013}. Lightning, which is closely associated with high updraft speeds \citep{Deierling2008}, further modulates atmospheric chemistry. Focusing on a basic property such as convective vigor, therefore, may reveal patterns in moist convective behavior with widespread implications. 

Currently, there are two paradigms for how the vigor of moist convection depends on thermodynamic conditions. The first paradigm, which we can call ``warming-driven invigoration'', states that moist convective vigor increases with the surface temperature. This paradigm has emerged from numerical modeling of Earth's contemporary tropics: in idealized cloud-resolving simulations, mean and extreme convective updraft speeds increase with warming \citep{Romps2011a,Muller2011,Seeley2015,Singh2015,Abbott2020}. These changes are consistent with, and typically attributed to, the increases in convective available potential energy (CAPE) that occur in both global and convection-resolving models \citep[e.g.,][]{DelGenio2007,Sobel2011a,Romps2011a,Muller2011,Singh2013,Seeley2015}. Although actual updrafts do not attain the velocities implied by CAPE due to drag and entrainment of unsaturated environmental air, CAPE is nevertheless an extremely useful proxy for intense convection \citep[e.g.,][]{Johns1992,Romps2014}. As shown by \cite{romps2016cape}, the increase of CAPE with warming under contemporary conditions is itself attributable to the Clausius-Clapeyron scaling of near-surface saturation specific humidity. Therefore, the chain of causality in the ``warming-driven invigoration'' paradigm is: 
$$\text{increasing } T_s \rightarrow \text{increasing humidity} \rightarrow \text{increasing CAPE} \rightarrow \text{increasing convective vigor}.$$

The second paradigm for convective vigor is relevant to atmospheres in which the condensible component is highly non-dilute, with the limiting case being a pure steam atmosphere. \cite{Ding2016} and \cite{Pierrehumbert2016} argued that convection should be gentle in such cases by reasoning that, when the atmosphere is saturated, the Clausius-Clapeyron curve dictates a one-to-one relationship between pressure and temperature, which precludes appreciable temperature anomalies between parcels at the same pressure level\footnote{This argument neglects ``virtual'' effects of compositional differences and condensates on density.}. These theoretical predictions gained support from the work of \cite{Tan2021}, who conducted idealized convection-resolving simulations of pure steam atmospheres with surface temperatures of 600--800 K and found quiescent condensing layers with small temperature variability and weak updrafts. Hence the line of reasoning in the ``gentle pure-steam limit'' is:
$$\text{very high humidity} \rightarrow \text{unique relationship between }T\text{and }p \rightarrow \text{low convective vigor}.$$

\noindent Clearly, ``warming-driven invigoration'', when extrapolated to very warm temperatures, conflicts with the ``gentle pure-steam limit''. Taken together, these paradigms suggest that convective vigor should reach a peak at some intermediate humidity level.

In this paper, we seek to generalize our understanding of moist convective vigor by filling the gap between these two paradigms. Our approach is to simulate a wide range of planetary atmospheres that differ appreciably from contemporary Earth in terms of their surface temperature, surface pressure, gravitational constant, and composition. These atmospheres span the parameter space from highly dilute to non-dilute conditions, allowing us to continuously probe the transitional behavior between the two paradigms for convective vigor. To circumvent the inherent uncertainty of convective parameterizations and allow detailed analysis of convective dynamics, we performed these simulations with a flexible convection-resolving model (described in detail in section \ref{sec:exp}). We analyze metrics of convective vigor in these atmospheres, and show that our results match the predictions of an analytical theory for radiative-convective equilibrium (RCE) originally developed in the context of Earth's tropics \citep{romps2016cape} (hereafter, R16). The analytical theory of R16 predicts the mean CAPE of a convecting atmosphere, and we show that this theory provides a unifying theoretical framework that links the two paradigms for convective vigor.

The outline of the paper is as follows. In section \ref{sec:exp}, we describe the flexible convection-resolving model and our numerical experiments. Section \ref{sec:results} analyzes convective vigor in the simulated atmospheres. In section \ref{sec:peak_exp}, we apply the theory of R16 to our results. We conclude in section \ref{sec:disc} with a discussion of the implications of our work for early Earth and other planetary atmospheres.

\section{Experimental methods}\label{sec:exp}

\subsection{Core experiments}

Our core suite of convection-resolving RCE simulations consists of three experiments. The first core experiment used the Earth-like model configuration, with the total surface pressure fixed at the contemporary value of 10$^5$ Pa and the surface temperature varying from 275 K to 365 K (experiment name \emph{EarthTemp}). In the second core Earth-like experiment (\emph{EarthPressure}), the surface temperature was instead fixed at 300 K while the surface pressure varied from (1/16)$\times$ to 8$\times$ the contemporary value. The third core experiment (\emph{Titan}) used the Titan-like configuration of the model with the surface pressure fixed at the contemporary value of 1.467$\times$10$^5$ Pa and the surface temperature varying from 80 to 110 K. The parameters distinguishing the Earth-like and Titan-like configurations will be described in detail below; see also Table \ref{tab:param}. A number of additional simulations were performed as sensitivity tests, which we will describe as they come up in the course of the main text.

\subsection{Convection-resolving model}

For all experiments, we simulated nonrotating radiative-convective equilibrium on square, doubly periodic domains with the convection-resolving model DAM \citep{Romps2008}. DAM has a finite-volume, fully-compressible dynamical core and uses the implicit approach to sub-grid diffusion. The vertical grid had a variable spacing, transitioning smoothly from $\Delta z=50$ m below an altitude of 500 m to $\Delta z=1000$ m at altitudes between 10 km and the model top. The model top was at a variable height because our simulated tropospheres vary widely in geometric depth. The default horizontal resolution was $\Delta x =\Delta y= 2$ km and the default horizontal domain size was $L_x=L_y=96$ km. We also ran a small subset of simulations at a higher resolution of $\Delta x = 500$ m and with free-tropospheric $\Delta z=500$ m (the \emph{EarthTemp\_hr} experiment). The default model time step was $\Delta t=20$~s, sub-stepped to satisfy a CFL condition; this time step was used for all simulations but \emph{EarthTemp\_hr}, which used $\Delta t=5$~s. Overall, the model domains are similar in size and resolution to the ``RCE\_small" protocol from the RCEMIP project \citep{wing2017b} in which DAM participated.

\subsection{Radiative transfer}

To avoid inessential complexity and focus attention on convective dynamics, we used a simplified treatment of radiative transfer in the majority of our simulations \citep[as in, e.g.,][]{Tan2021}. Specifically, we prescribed an idealized tropospheric radiative cooling using an equation of the form:

\begin{equation}
    -\frac{\partial}{\partial T} F = \alpha (T - T_t), \label{eq:dT_F}
\end{equation}

\noindent where $T$ is the temperature, the temperature derivative $\partial/\partial T$ is a vertical derivative, $\alpha$ (W/m$^2$/K$^2$) is a constant setting the magnitude of radiative forcing, $F$ (W/m$^2$) is the net upward radiative flux, and $T_t$ (K) is the prescribed tropopause temperature. Numerically, we approximated the vertical derivative $\partial/\partial T$ with a centered finite difference on the model's vertical grid. Equation \ref{eq:dT_F} was proposed by \cite{Jeevanjee2018b} as a fit to the invariant radiative flux divergence curve found for Earth, and has been used as an idealized representation of tropospheric radiative cooling in more recent work \citep{Jeevanjee2022a}. We used equation (\ref{eq:dT_F}) to prescribe radiative forcing for both Earth-like and Titan-like experiments, although we used different values of the parameters (Table \ref{tab:param}). At altitudes above the tropopause, temperatures were simply nudged to $T_t$ on a timescale of 6 hours; since our focus is on convective dynamics in the troposphere, this simplified approach to the stratosphere does not affect our results.

One deficiency of equation (\ref{eq:dT_F}) is that it does not predict the transition to lower-tropospheric radiative \emph{heating} in very warm climates, a regime which is known to affect convective dynamics \citep{Seeley2021}. Therefore, we also re-ran the \emph{EarthTemp} experiment with interactive clear-sky radiative transfer (the \emph{EarthTemp\_realrad} experiment) to check that our main results are not sensitive to the simplified treatment of radiation we employ in our core experiments. We focused on interactive clear-sky radiation because that is sufficient to produce the low-level radiative heating regime of \cite{Seeley2021}, and because DAM is not coupled to a cloud-radiation scheme that is validated up to the very high temperatures we simulate. Our approach to clear-sky radiative transfer was identical to the line-by-line method of \cite{Seeley2021}, and we refer the reader to that paper for a complete description.

\subsection{Microphysics}

The default microphysics scheme in DAM is a bulk scheme with six water classes (vapor, cloud liquid, cloud ice, rain, snow, and graupel). However, to avoid relying on an overly Earth-centric parameterization, we conducted our simulations using a simplified microphysics scheme that has been described and used in previous work \citep{Seeley2021}. In the simplified microphysics scheme, only three bulk classes of condensible substance are modeled: vapor, non-precipitating cloud liquid, and rain, with associated mass fractions $q_v$, $q_c$, and $q_r$, respectively. Microphysical transformations between vapor and cloud condensate are handled by a saturation adjustment routine, which prevents relative humidity from exceeding 100\% (i.e., abundant cloud condensation nuclei are assumed to be present) and evaporates cloud condensate in subsaturated air. Conversion of non-precipitating cloud condensate to rain is modeled as autoconversion according to
\begin{equation}
a = -q_c/\tau_a,
\end{equation}
\noindent where $a$ (s$^{-1}$) is the sink of cloud condensate from autoconversion and $\tau_a$ (s) is an autoconversion timescale. We use $\tau_a=25$ minutes, which was found in prior work to produce a similar mean cloud fraction profile as the default bulk scheme in DAM \citep{Seeley2020}. We did not set an autoconversion threshold for $q_c$. Rain is given a fixed freefall speed with a default value of 8 m/s, but we also checked the sensitivity of our results to this value when appropriate (the \emph{VaryGrav\_fs} experiment). When rain falls through subsaturated air, it is allowed to evaporate according to

\begin{equation}
e = (q_v^* - q_v)/\tau_r,
\end{equation}
\noindent where $e$ (s$^{-1}$) is the rate of rain evaporation, $q_v^*$ is the saturation specific humidity, and $\tau_r$ (s) is a rain-evaporation timescale. We set $\tau_r=50$ hours, which was found in prior work to produce a tropospheric relative humidity profile similar to that of the bulk scheme \citep{Seeley2020}. Since microphysics on Titan is very poorly constrained, we used the same values for these microphysical constants in both the Earth-like and Titan-like model configurations; future work could attempt to use first-principles theories to tune microphysical parameters for the Titan regime \citep{lorenz1993life,Loftus2021}.

\begin{table}[]
\centering
\begin{tabular}{@{}|c|l|l|l|@{}}
\hline
\textbf{parameter}  & \textbf{description}                                                                    & \textbf{Earth-like}          & \textbf{Titan-like}        \\ \hline
$g$        & gravitational constant (m/s$^2$)                                               & 10.                  & 1.35              \\
$T_t$      & tropopause temperature (K)                                                     & 200.                 & 70.                \\
$\alpha$   & radiative forcing constant (W/m$^2$/K)                                         & 0.025               & 0.5               \\
$R_a$      & specific gas constant of dry air (J/kg/K)                                      & 287.04              & 296.8             \\
$R_v$      & specific gas constant of condensible gas (J/kg/K)                              & 461.                & 518.28            \\
$c_{va}$   & specific heat capacity at constant volume of dry air (J/kg/K)                  & 719.                & 707.2             \\
$c_{vv}$   & specific heat capacity at constant volume of condensible gas (J/kg/K)          & 1418.               & 1707.4            \\
$c_{vl}$   & specific heat capacity at constant volume of condensible liquid (J/kg/K)       & 4119.               & 3381.55           \\
$c_{pv}$   & specific heat capacity at constant pressure of condensible gas (J/kg/K)        & $c_{vv}+R_v$        & $c_{vv}+R_v$      \\
$E_{0v}$   & internal energy difference between vapor and liquid at the triple point (J/kg) & 2.374$\times$10$^6$ & 4.9$\times$10$^5$ \\
$p_{trip}$ & pressure at condensible's triple point (Pa)                                      & 611.65              & 11700.            \\
$T_{trip}$ & temperature at condensible's triple point (K)                                    & 273.16              & 90.68             \\ \hline
\end{tabular}
\caption{Parameters for the convection-resolving experiments in the Earth-like and Titan-like model configurations.}\label{tab:param}
\end{table}

\subsection{Thermodynamics}

The principal difference between our Earth-like and Titan-like model configurations pertains to the atmospheric thermodynamics. The thermodynamics of moist air in DAM is based on a standard set of approximations applying to mixtures of dry air and a condensible component which may be present in vapor, liquid, and solid form. These approximations are 1) both dry air and the condensible vapor are treated as ideal gases, 2) the heat capacities of all components are assumed not to depend on temperature, and 3) condensates are assumed to have zero specific volume \citep{Ambaum2010,Romps2008,Romps2021}.

In our convection-resolving experiments, we use a simplified treatment of the condensible component by neglecting the solid phase, which is not of leading importance for convective dynamics in Earth's tropics \citep{Seeley2016}. Therefore, at all temperatures saturation is  evaluated via the expression for the saturation vapor pressure over liquid, which is derived from the above approximations and implemented in DAM as

\begin{equation}
    p_v^* = p_{trip}\left(\frac{T}{T_{trip}}\right)^{(c_{pv}-c_{vl})/R_v} \exp\left(\frac{L_e(T_{trip})}{R_v T_{trip}} - \frac{L_e(T)}{R_v T}\right), \label{eq:pvstar}
\end{equation}
\noindent where $L_e$ is the temperature-dependent latent enthalpy of evaporation given by
\begin{equation}
    L_e(T) = E_{0v} + R_v T + (c_{vv}-c_{vl})(T-T_{trip}), \label{eq:Le}
\end{equation}
\noindent and where the other physical constants appearing in equations (\ref{eq:pvstar}--\ref{eq:Le}) are defined in Table \ref{tab:param}. 

The values assigned to these and other physical constants determine whether DAM simulates Earth-like or Titan-like atmospheric composition and moist thermodynamics (Table \ref{tab:param}). For our Earth-like configuration, dry air is assumed to be a mixture of 80\% N$_2$ and 20\% O$_2$, while the condensible component is water (H$_2$O). For the Titan-like configuration, dry air is assumed to be entirely N$_2$ and the condensible component is methane (CH$_4$).

\subsection{Surface fluxes}

Surface fluxes were modeled with bulk aerodynamic formulae. Specifically, the surface latent and sensible heat fluxes (LHF and SHF) were given by 

\begin{equation}
    \mathrm{LHF}(x,y) = \rho_1(x,y) C_D\sqrt{u_1(x,y)^2 + v_1(x,y)^2 + V^2} L_e  \left[ q_s^* - q_1(x,y)\right];
\end{equation}

\begin{equation}
    \mathrm{SHF}(x,y) = \rho_1(x,y) C_D\sqrt{u_1(x,y)^2 + v_1(x,y)^2 + V^2} c_p  \left[ \mathrm{SST} - T_1(x,y)\right],
\end{equation}
\noindent where $\rho_1$, $q_1$, $u_1$, $v_1$, and $T_1$ are the density, specific humidity, horizontal winds, and temperature at the first model level, $C_D=1.5\times10^{-3}$ is a drag coefficient, $V=5$ m/s is a background ``gustiness'', $L_e$ is the latent heat of evaporation, $c_p$ is the specific heat capacity at constant pressure of moist air, and $q_{s}^*$ is the saturation specific humidity at the sea surface temperature and surface pressure. Since the time-mean surface enthalpy flux is constrained by the (imposed) tropospheric radiative cooling, the values of $C_D$ and $V$ determine the near-surface air-sea enthalpy disequilibrium but do not otherwise affect our results.

\section{Results}\label{sec:results}

\begin{figure}[ht!]
\centerline{\includegraphics[width=\textwidth]{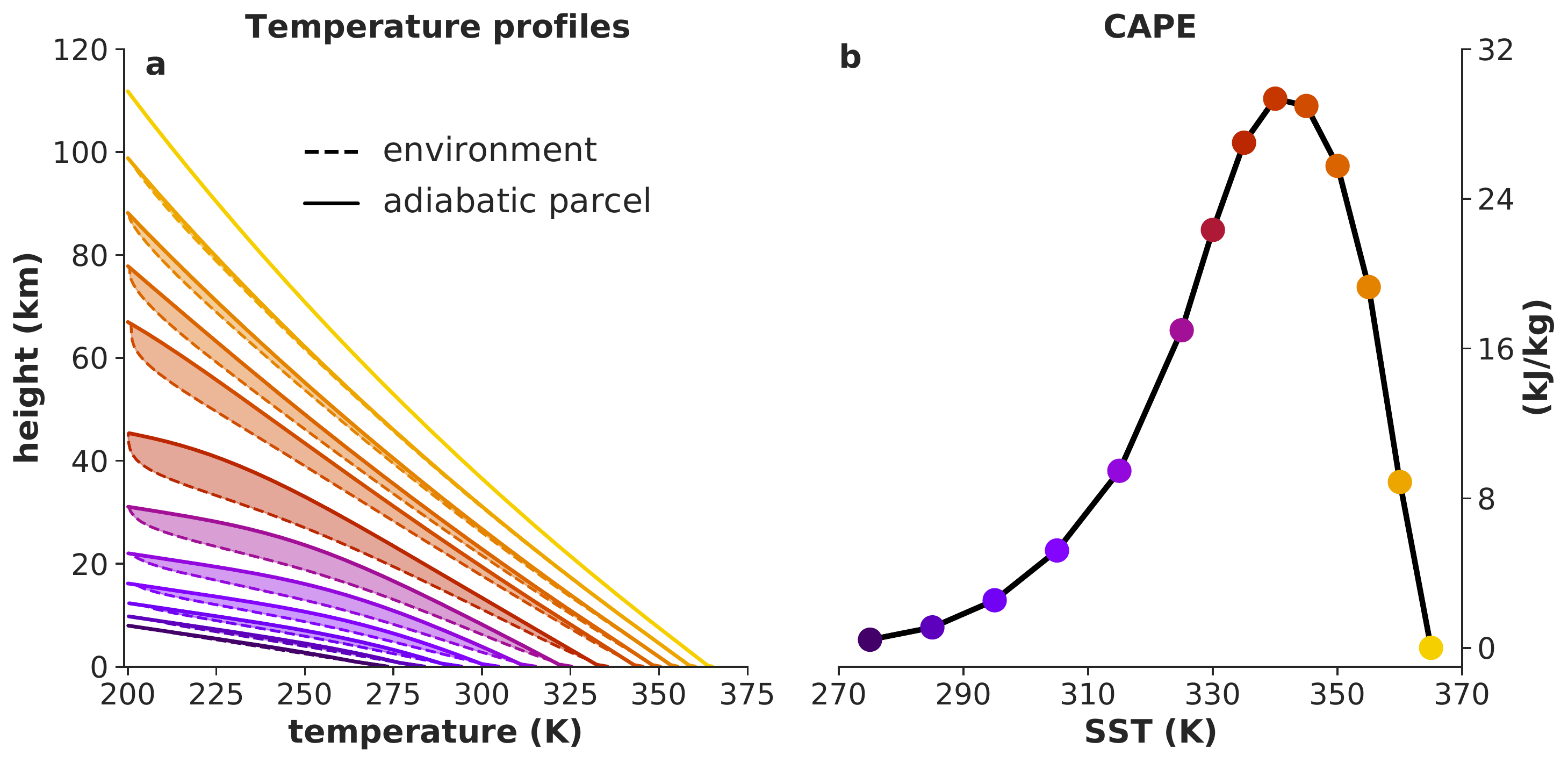}}
\caption{(a) Temperature profiles of the environment (dashed) and adiabatically-lifted near-surface parcels (solid) from the \emph{EarthTemp} experiment; the area between these two profiles is shaded where the parcel is warmer than the environment. For visual clarity, a subset of surface temperature cases are plotted, and the environmental temperature profile is only plotted where it is cooler than the adiabatic parcel. (b) Convective available potential energy (CAPE), defined as the vertically-integrated positive buoyancy of the lifted parcels whose temperature profiles are shown in panel (a). The condensate mass fraction in the lifted parcels was subjected to an exponential-decay sink term with a vertical length scale of $L=5$ km; see Figure \ref{fig:fallout} for the effect of different condensate fallout assumptions.}
\label{fig:vary_Ts}
\end{figure}

We begin our study of convective vigor with a focus on CAPE. CAPE is the maximum specific vertical kinetic energy, $w^2/2$ (where $w$ is vertical velocity), that clouds can attain while rising, so it is a useful summary statistic for the potential vigor of convection. Here we calculate CAPE as the vertically-integrated positive buoyancy $b$ of an adiabatically-lifted parcel between its lifted condensation level (LCL) and its level of neutral buoyancy (LNB):

\begin{equation}
    \mathrm{CAPE} = \int_\mathrm{LCL}^\mathrm{LNB} \mathrm{max}(b,0)\ dz, \label{eq:CAPE}
\end{equation}
\noindent where the buoyancy of the parcel is  $b = g(\rho_e/\rho_p -1)$ for parcel density $\rho_p$ and environmental density $\rho_e$.

We first examine CAPE in the \emph{EarthTemp} experiment, for which the surface temperature was varied in the Earth-like model configuration. Figure \ref{fig:vary_Ts}a shows, for a subset of surface temperature cases, the environmental temperature profile (time- and horizontal-mean) compared to the temperature profile of an adiabatically-lifted parcel that is initialized with the thermodynamic properties of mean near-surface air. There is a clear pattern in these temperature profiles: at low and high surface temperatures, the environment and the adiabat\footnote{For brevity, we will refer to the temperature profile of a lifted, undiluted near-surface parcel as an ``adiabat'', ignoring the small effect of buoyancy on the parcel's lapse rate \citep[e.g.,][]{Riehl1958,Romps2015b}. Note that the ``adiabats'' plotted in Figure \ref{fig:vary_Ts} are intermediate between the pseudo-adiabatic process (all condensates removed) and the reversible process (all condensates retained). Specifically, the condensate sink term from fallout is modeled as $\frac{\partial q_c}{\partial z}|_{fall} = -q_c/L$} are nearly indistinguishable, whereas for intermediate surface temperatures the adiabat is significantly warmer than the environment, especially in the upper troposphere. We will describe the physical explanation for this behavior in section \ref{sec:peak_exp}. Since CAPE measures the integrated buoyancy of an adiabatic parcel, CAPE is therefore small at both low and high surface temperatures, with a peak in between (Fig. \ref{fig:vary_Ts}b). While the quantitative magnitude and location of this peak in CAPE are somewhat sensitive to assumptions about condensate fallout in the lifted adiabatic parcel (Fig. \ref{fig:fallout}), the existence of a peak is robust to these details.

\begin{figure}[ht!]
\centerline{\includegraphics[width=\textwidth]{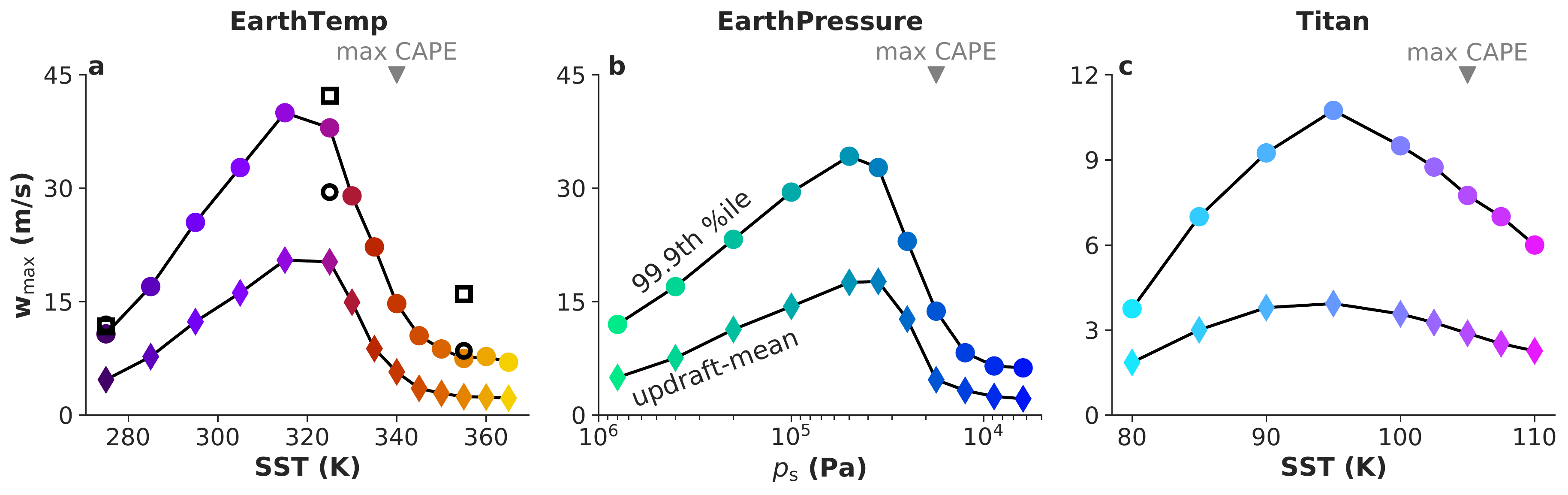}}
\caption{Metrics of actual convective vigor for the (a) \emph{EarthTemp}, (b) \emph{EarthPressure}, and (c) \emph{Titan} experiments. The metric $w_\mathrm{max}$ refers to the maximum tropospheric value of the profile of mean updraft velocity (diamond markers) or the profile of 99.9th percentile of vertical velocity at each altitude (circle markers). In each panel, the location on the $x$-axis with maximum CAPE is marked by the triangle at the top of the plot}. In panel (a), the black open circles and black open squares show $w^{99.9}_\mathrm{max}$ from the simulations at 275, 325, and 355 K with high resolution (\emph{EarthTemp\_hr}) and with interactive clear-sky radiation (\emph{EarthTemp\_realrad}), respectively. Note that panel (c) has a different $y$-axis limit than panels (a) and (b).
\label{fig:updraft}
\end{figure}

The growth and decline of CAPE with warming in the \emph{EarthTemp} experiment appears to connect the previously mentioned ``warming-driven invigoration'' and ``gentle pure-steam limit'' regimes. The near-surface specific humidity in these simulations ranges from about 0.2\% in the coldest simulation to about 60\% in the warmest, confirming that water vapor transitions from being a very minor trace gas to a dominant component. However, CAPE only measures the \emph{potential} vigor of convection. How do actual convective updraft speeds vary in these simulations? Figure \ref{fig:updraft}a shows two metrics of actual convective vigor from the \emph{EarthTemp} experiment. The first of these metrics,  $w_\mathrm{max}^\mathrm{mean}$, is calculated by conditionally sampling all grid cells with $w>1$ m/s and cloud condensate $q_c>10^{-5}$ kg/kg. The horizontal-mean vertical velocity of these cloudy-updraft grid cells is calculated at each tropospheric vertical grid level, and the maximum value of this profile is what we refer to as $w_\mathrm{max}^\mathrm{mean}$ (labeled ``updraft-mean'' in Fig. \ref{fig:updraft}b). The second metric of convective vigor is calculated from the histogram of vertical velocities at each tropospheric vertical grid level; these histograms are not conditionally sampled based on cloud condensate. We calculate the 99.9th percentile of each tropospheric grid level's vertical velocity histogram, and the maximum value of this vertical profile is what we refer to as $w_\mathrm{max}^{99.9}$ (labeled ``99.9th percentile'' in Fig. \ref{fig:updraft}b). Similar to CAPE, these metrics also show an initial growth and eventual decline with increasing surface temperature, with the warmest simulation actually having slower $w_\mathrm{max}$ values than the coldest. Note that although the qualitative behavior of $w_\mathrm{max}$ is similar to CAPE, the peaks in the $w_\mathrm{max}$ metrics occur at a lower surface temperature than the CAPE peak, and $w_\mathrm{max}$ is right-skewed while CAPE is left-skewed. Other choices of summary statistics for convective vigor lead to peaks at slightly different surface temperatures, but the overall phenomenon is robust to these choices. The peak in convective vigor is also robust to increased horizontal and vertical resolution (the \emph{EarthTemp\_hr} experiment; Fig. \ref{fig:updraft}a, open circles) and use of realistic clear-sky radiation (the \emph{EarthTemp\_realrad} experiment; Fig. \ref{fig:updraft}a, open squares).

\begin{figure}[ht!]
\centerline{\includegraphics[width=\textwidth]{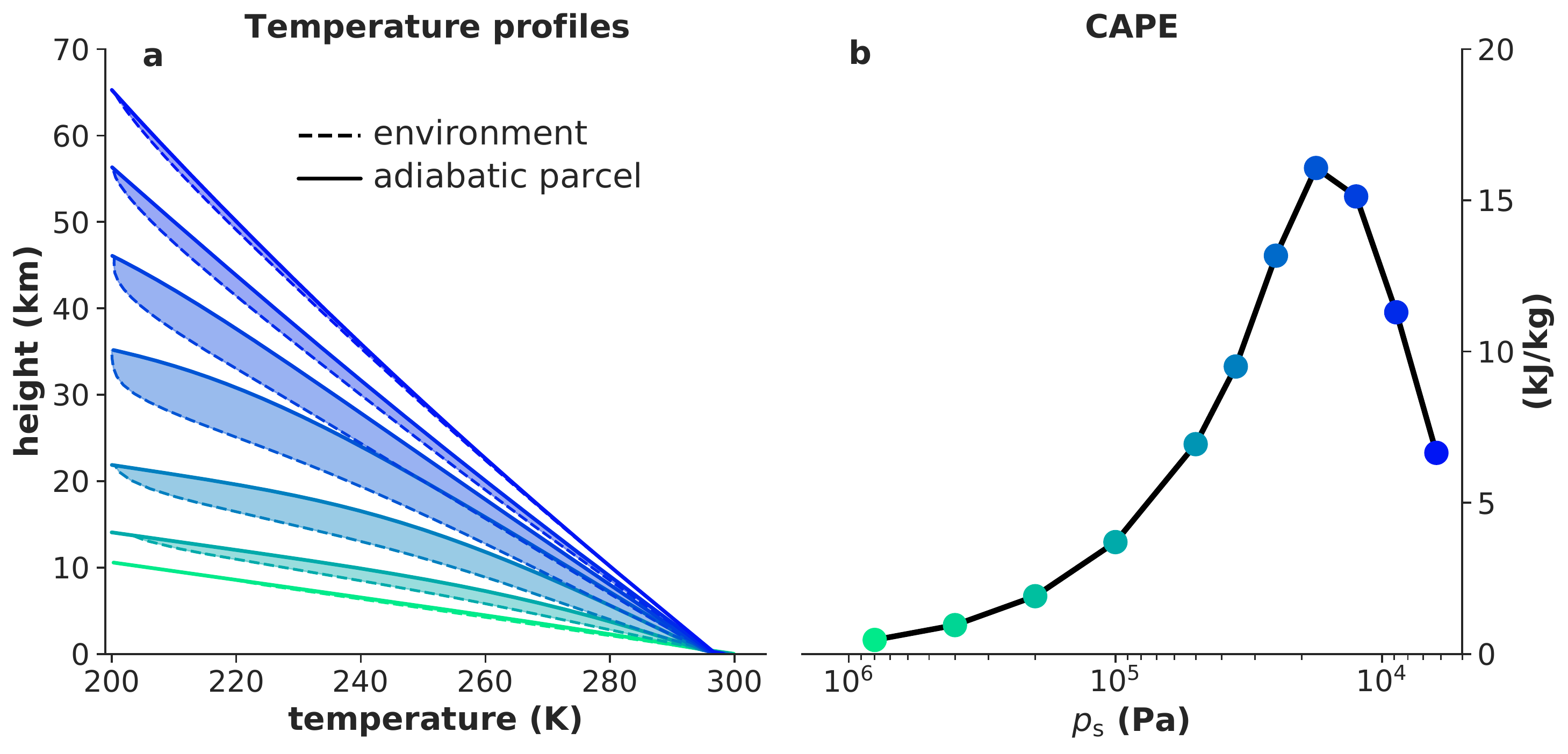}}
\caption{As in Figure \ref{fig:vary_Ts}, but for the \emph{EarthPressure} experiment. Note that in (b) the horizontal axis is inverted so that specific humidity increases toward the right.}
\label{fig:vary_ps}
\end{figure}

How general is this peak in convective vigor with respect to atmospheric humidity? In the \emph{EarthTemp} experiment, the increase in $q_v^*$ is driven by the increasing surface temperature and associated Clausius-Clapeyron scaling of the saturation vapor pressure. However, specific humidity can also be increased, at fixed temperature, by lowering the amount of non-condensing background gas \citep{Wordsworth2013}. Would varying the surface pressure, therefore, also produce variations in CAPE and convective vigor? To test this, we turn to the \emph{EarthPressure} experiment, in which we fixed the surface temperature at 300 K but varied the surface pressure from 8$\times$10$^5$ Pa to 6250 Pa (between a factor of 8$\times$ and 1/16$\times$ the contemporary value). Figure \ref{fig:vary_ps} shows that CAPE varies in \emph{EarthPressure} in qualitatively the same manner as in \emph{EarthTemp}, reaching a peak at an intermediate surface pressure (around 2$\times$10$^4$ Pa, approximately 20\% of the contemporary value). Likewise, Figure \ref{fig:updraft}b shows that actual convective vigor in the \emph{EarthPressure} experiment also peaks at intermediate surface pressures, although there is again an offset between the peak CAPE and the peak in actual convective vigor. Since Earth's surface pressure is relatively unconstrained during the Hadean and Archean \citep{Kavanagh2015,som2016earth}, these results may have implications for the early evolution of Earth's climate and atmospheric chemistry; we will return to this topic in section \ref{sec:disc}.

\begin{figure}[ht!]
\centerline{\includegraphics[width=\textwidth]{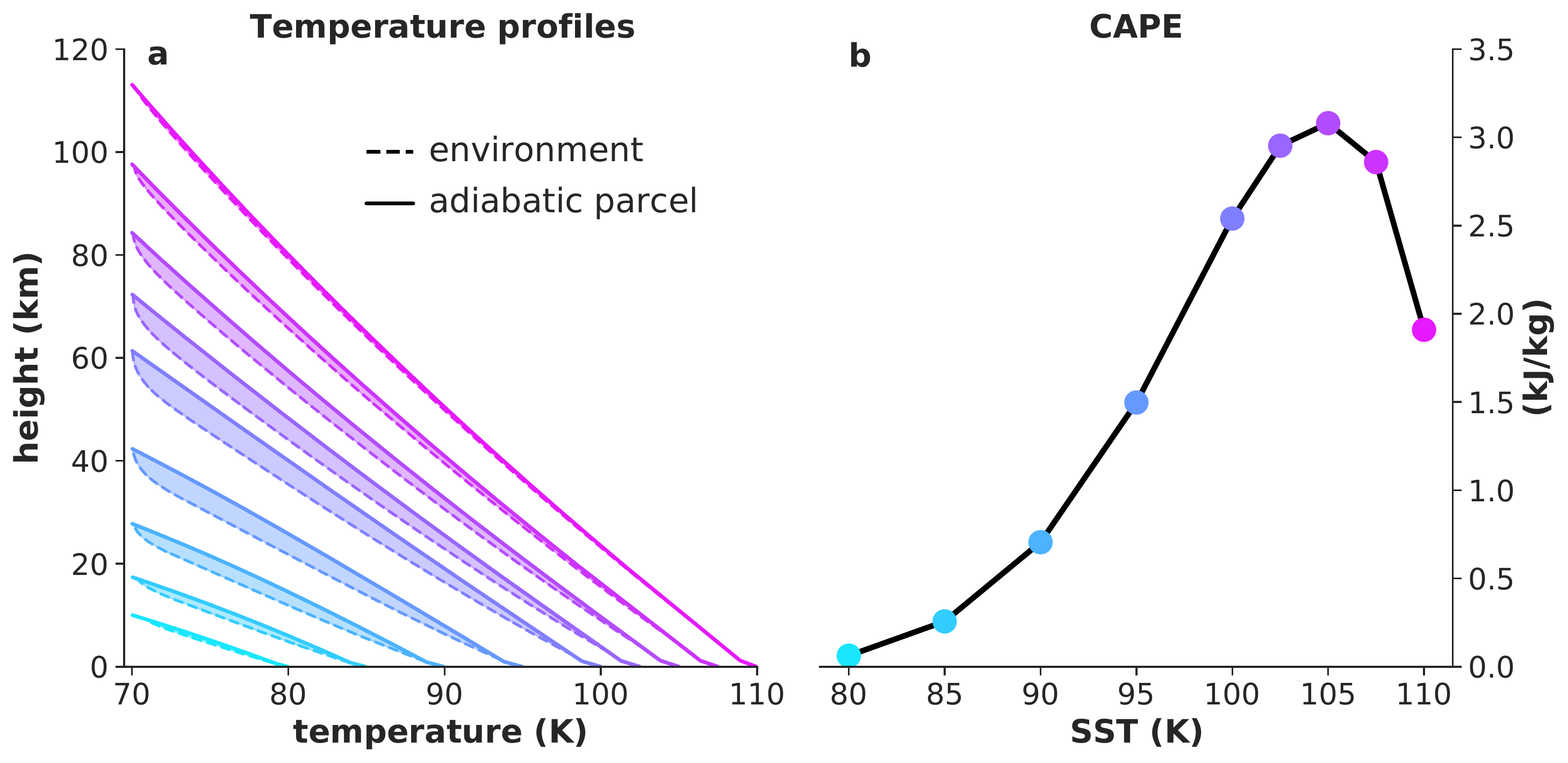}}
\caption{As in Figures \ref{fig:vary_Ts} and \ref{fig:vary_ps}, but for the \emph{Titan} experiment.}
\label{fig:titan}
\end{figure}

If the boom-bust evolution of CAPE seen in the \emph{EarthTemp} and \emph{EarthPressure} experiments is attributable to the transition from condensible-poor (dilute) to condensible-rich (non-dilute) conditions, this recipe is not specific to Earth-like moist convection: in any atmosphere with a sufficiently large surface reservoir of a condensible species, the condensible will become increasingly volatile with warming and eventually come to dominate atmospheric composition. To what other planetary atmospheres, then, might this boom-bust CAPE behavior apply? A natural candidate is Saturn's moon Titan, which is often regarded as the closest current hydrological analog to Earth due to its active methane precipitation cycle. Therefore, to further generalize our understanding of convective vigor, we next turn to the \emph{Titan} experiment. This experiment assumes Titan-like thermodynamic conditions and atmospheric composition (i.e., a thick N$_2$ atmosphere with condensing CH$_4$ and low gravity; Table \ref{tab:param}). Similar to the Earth-like experiments, Figure \ref{fig:titan} shows a peak in CAPE as the simulated Titan-like atmospheres transition from dilute to non-dilute methane abundance. Figure \ref{fig:updraft}c shows that metrics of actual convective vigor in this experiment peak at a surface temperature of about 95 K. Intriguingly, this is very close to the current surface temperature of Titan.

\begin{figure}[ht!]
\centerline{\includegraphics[width=\textwidth]{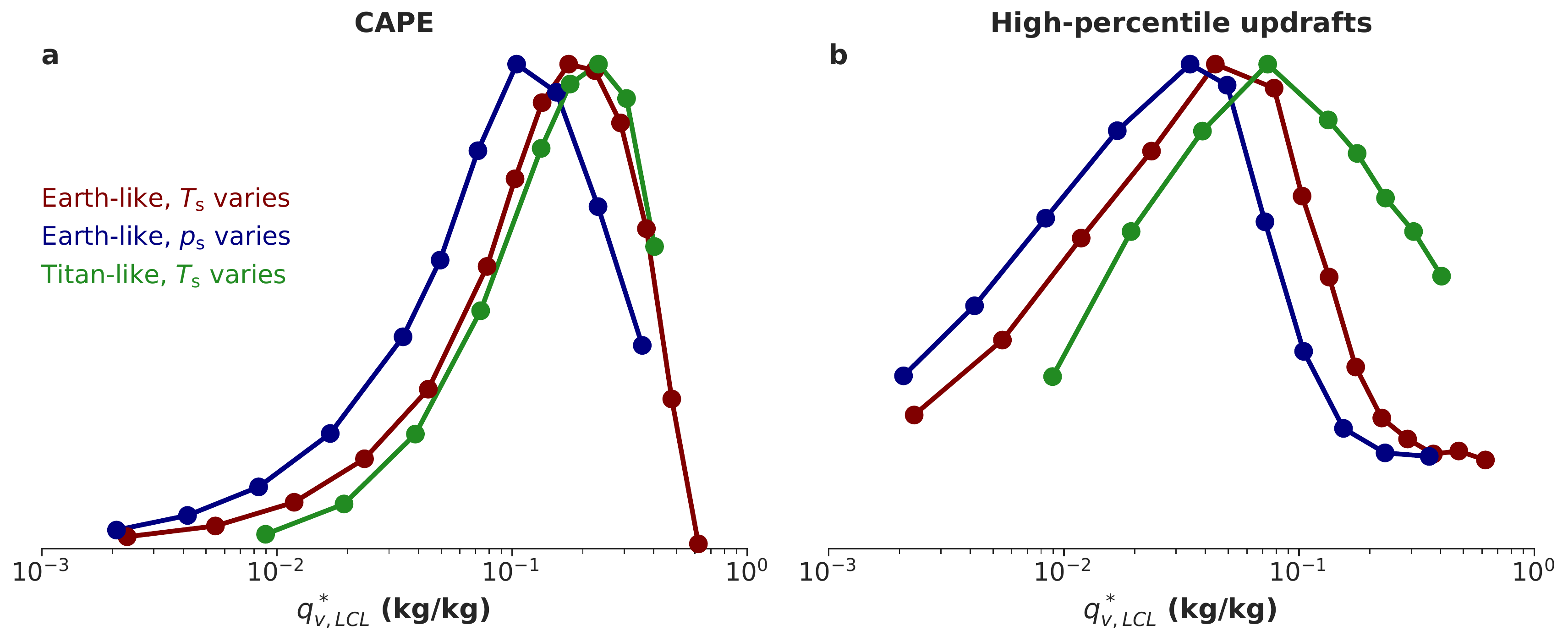}}
\caption{(a) Normalized CAPE from the \emph{EarthTemp}, \emph{EarthPressure}, and \emph{Titan} experiments, plotted as a function of the saturation specific humidity at the lifted condensation level $q_{v,LCL}^*$. (b) As in (a), but for normalized high-percentile updraft speeds ($w_{max}^{99.9}$).}
\label{fig:qLCL}
\end{figure}

To better compare our three core experiments, it is helpful to recast the results in terms of variations in atmospheric humidity rather than surface temperature or pressure. Figure \ref{fig:qLCL} plots CAPE and high-percentile updraft speeds from the core experiments as a function of the specific humidity at the lifted condensation level, $q_{v,LCL}^*$. This reveals that CAPE and convective vigor in all three experiments peaks when cloud base air contains roughly 10\% of the condensible component by mass, give or take a factor of about 2. Therefore, we can conclude that the ``warming-driven invigoration'' regime comes to an end at intermediate humidity, well before these atmospheres approach the steam limit.

\section{The physical origin of the CAPE peak}\label{sec:peak_exp}

Taken all together, our three core experiments point to a potentially common phenomenon in terrestrial atmospheres: moist convection is most vigorous at intermediate atmospheric humidity. What is the physical explanation for this behavior? In this section, we show that recent advances in the theory of convection provide a quantitative explanation for the peak in CAPE with respect to atmospheric humidity.

The basic ingredient required to understand climatological variations in CAPE is a theory for the tropospheric lapse rate, which we denote by $\Gamma(z) = -\partial T/\partial z$ (K/km). Note that the standard assumption in idealized 1-dimensional radiative-convective modeling, which is that $\Gamma(z)$ is given by the moist adiabat, is useless for the purpose of predicting CAPE: the CAPE of a moist-adiabatic atmosphere is zero by definition. It is the systematic \emph{deviations} from a moist-adiabatic thermal structure that a successful theory for CAPE must predict. 

The key theoretical breakthrough in this regard was made by \cite{Singh2013} (hereafter, SO13), who emphasized the role of entrainment. Entrainment refers to the turbulent mixing with environmental air that cloudy updrafts experience as they ascend. Because entrainment of subsaturated air reduces condensation and latent heating in ascending parcels, entraining parcels cool more rapidly with height than otherwise identical undiluted parcels. The insight of SO13 is that the troposphere can be approximated as being neutrally buoyant with respect to such \emph{entraining} convective parcels, rather than with respect to an undiluted parcel; this assumption has come to be known as the ``zero-buoyancy" (ZB) approximation\footnote{We stress that this zero-buoyancy assumption is not a zero-CAPE assumption: it is the \emph{entraining} convective parcels that are assumed to have zero buoyancy with respect to the mean environment, not the adiabatic parcel that is used to calculate CAPE.}. Defining $\Gamma_m$ as the lapse rate of an undiluted parcel and $\Gamma_e$ as the lapse rate of an entraining parcel (which, by the ZB approximation, is equal to $\Gamma$), we can state that $\Gamma_m < \Gamma = \Gamma_e$. According to this picture, then, entrainment is the wedge that drives $\Gamma_m$ and $\Gamma$ apart, allowing for potentially large reservoirs of CAPE even in steady-state RCE.


To make this discussion quantitative, we turn to the simplest model of a convecting atmosphere that incorporates entrainment: the ``bulk-plume model"\footnote{This discussion of the zero-buoyancy bulk-plume theory for CAPE is based on that in \cite{romps2016cape} and \cite{Romps2020a}; we refer the reader to these references for a more thorough treatment.}. The bulk-plume model divides the atmosphere into two plumes: ascending, saturated (cloudy) air, and descending, subsaturated environmental air; the ``bulk" descriptor refers to the fact that the thermodynamic properties of these two plumes are assumed to be homogeneous in the horizontal at each altitude. Mass exchange between the two plumes is represented by specified entrainment and detrainment rates, such that conservation of mass in the bulk-plume model is expressed as
\begin{eqnarray}
\frac{\partial M}{\partial z} &=& e - d \nonumber \\
&=& M(\varepsilon - \delta), \label{eq:masscon}
\end{eqnarray}
\noindent where $M$ (kg/m$^2$/s) is the upward convective mass flux (equal and opposite, in RCE, to the subsidence mass flux in the environmental plume), $e$ and $d$ (kg/m$^3$/s) are the mass entrainment and detrainment rates, and $\varepsilon$ and $\delta$ (m$^{-1}$) are known as the fractional entrainment and detrainment rates, defined as $e/M$ and $d/M$, respectively. Equation (\ref{eq:masscon}) implies that the convective mass flux increases with height if entrainment outpaces detrainment, and vice versa. 

The second bulk-plume equation we will use is the conservation equation for moist static energy $h$, which is conventionally defined as $h=c_p T + g z + L q_v$. Here $c_p$ (J/kg/K) is the specific heat capacity of air at constant pressure, $L$ (J/kg) is the latent heat of evaporation, and the other symbols take their usual meaning. While this is an approximate expression for moist static energy --- neglecting, for instance, the temperature-dependence of the latent heat of evaporation \citep{Romps2015b} --- it is sufficiently accurate to form the basis of a theory for CAPE. The conservation of moist static energy in the entraining convective plume is expressed as
\begin{equation}
\frac{\partial (M h^*)}{\partial z} = e h_e - dh^*, \label{eq:msecon}
\end{equation}
\noindent where $h^*=c_p T + g z + L q_v^*$ is the saturation moist static energy (appropriate for the convective plume because it is saturated). The environmental moist static energy can be expressed as $h_e = c_p T + g z + L \mathcal{R} q_v^*$, where we have used the same $T$ as in the convective plume (invoking the ZB approximation) and where the environmental specific humidity is given by the product of $\mathcal{R}$, the environmental relative humidity, and the saturation specific humidity, $q_v^*$.

Using the product rule on the left-hand side of equation (\ref{eq:msecon}), substituting in equation (\ref{eq:masscon}), and using the definitions of $h^*$ and $h_e$ given above, we arrive at an important result from SO13:
\begin{equation}
\frac{\partial h^*}{\partial z} = - \varepsilon\left(1-\mathcal{R}\right)L q_v^*, \label{eq:satdef}
\end{equation}
\noindent Equation \ref{eq:satdef} states that the entraining plume's moist static energy decreases with height at a rate proportional to the saturation deficit of the environment, $(1-\mathcal{R})q_v^*$ \citep{Seeley2015}.

Although SO13 treated the environmental relative humidity, $\mathcal{R}$, as an external parameter that must be specified, analysis of the bulk-plume water budget can yield a self-consistent analytic expression for $\mathcal{R}$ \citep{Romps2014a}:
\begin{equation}
\mathcal{R} = \frac{\delta + \alpha \gamma - \alpha \varepsilon}{\delta + \gamma - \alpha \varepsilon}.\label{eq:RH}
\end{equation}
\noindent In equation \ref{eq:RH}, $\alpha$ is a nondimensional parameter specifying the fraction of condensates formed at a given height that evaporate at that height instead of precipitating out of the atmosphere (i.e., the \emph{precipitation efficiency} of the bulk-plume convection is $1-\alpha$). The quantity $\gamma$ is the water vapor lapse rate, defined as $\gamma \equiv - \partial \ln q_v^*/\partial z$ and expressed in terms of thermodynamic parameters as
\begin{equation}
\gamma = \frac{L \Gamma}{R_v T^2} - \frac{g}{R T},\label{eq:gammawater}
\end{equation}
\noindent where $R_v$ and $R$ (J/kg/K) are the specific gas constants for water vapor and dry air, respectively. The expression (\ref{eq:gammawater}) is straightforward to derive by combining the Clausius-Clapeyron equation for the saturation vapor pressure with hydrostatic balance \citep{Romps2014a}.

The final step toward the theory for CAPE was taken by \cite{romps2016cape}. For analytic solubility, that work considered a simplified case and assumed that $M$, $\mathcal{R}$, and $\alpha$ are all constant with height. The constancy of $M$ implies $\varepsilon=\delta$, by equation (\ref{eq:masscon}), while the constancy of $\mathcal{R}$ and $\alpha$ imply that the relative humidity and entrainment rate can be expressed in terms of another constant, $a \geq 0$,  as follows:
\begin{eqnarray}
\mathcal{R} &=& \frac{a + \alpha}{1 + a}, \label{eq:RH2}\\
\varepsilon &=& a\left(\frac{\gamma}{1 - \alpha}\right) \label{eq:ent}.
\end{eqnarray}
\noindent Note that $a$, which we will refer to as the ``bulk-plume parameter'', is proportional to the entrainment rate, so that $a=0$ corresponds to undiluted convection. Plugging in equations (\ref{eq:RH2}--\ref{eq:ent}) to the right-hand side of equation (\ref{eq:satdef}) yields
\begin{equation}
\frac{\partial h^*}{\partial z} = -\frac{a}{1+a}\gamma L q_v^* \label{eq:dhdz}.
\end{equation}
\noindent We can obtain a second equation for $\partial h^*/\partial z$ by differentiating the expression for $h^*$ directly:
\begin{eqnarray}
\frac{\partial h^*}{\partial z} &=& -c_p \Gamma + g - \gamma L q_v^* \nonumber \\
&=& g\left(1 + \frac{q_v^* L}{R T}\right) + \Gamma \left(c_p + \frac{q_v^* L^2}{R_v T^2}\right), \label{eq:dhdz2}
\end{eqnarray}
\noindent where the second line follows from substituting equation (\ref{eq:gammawater}) for $\gamma$. Equating the right-hand sides of equations (\ref{eq:dhdz}) and (\ref{eq:dhdz2}) and solving for $\Gamma$, we obtain
\begin{equation}
\Gamma = \left(\frac{g}{c_p}\right) \left[\frac{1 + a + q_v^* L/(R T)}{1 + a + q_v^* L^2/(c_p R_v T^2)}\right]. \label{eq:gamma_ent}
\end{equation}

Equation (\ref{eq:gamma_ent}) is a generalization of the moist lapse rate that accounts for the effect of entrainment \citep{Romps2020a}. In the limit of no entrainment ($a\rightarrow0$), equation \ref{eq:gamma_ent} reduces to the standard expression for the moist adiabat, $\Gamma_m$:
\begin{equation}
\Gamma_m = \left(\frac{g}{c_p}\right) \left[\frac{1 + q_v^* L/(R T)}{1 + q_v^* L^2/(c_p R_v T^2)}\right]. \label{eq:gamma_m}
\end{equation}

We can use equation (\ref{eq:gamma_ent}) to analyse the difference between $\Gamma$ and $\Gamma_m$ --- and, therefore, CAPE --- in the limit of very dry and very moist atmospheres. In the dry limit ($q_v^* \rightarrow 0$), which is approached by reducing the surface temperature or increasing the amount of background dry air, we can drop all terms multiplied by $q_v^*$ inside the brackets in equation (\ref{eq:gamma_ent}), which allows the factors of $(1 + a$) in the numerator and denominator to cancel and yields $\Gamma_m = g/c_p = \Gamma_d$, the dry-adiabatic lapse rate. Hence entrainment has no effect on the lapse rate in the dry limit, $\Gamma$ and $\Gamma_m$ are both equal to the dry adiabat, and CAPE is zero. On the other hand, in the very moist limit ($q_v^* \rightarrow 1$), which is approached by increasing the surface temperature or decreasing the amount of background dry air, the terms multiplied by $q_v^*$ dominate over the factors of $a$ for typical Earth and Titan-like conditions. For example, for Earth-like conditions\footnote{We leave investigation of whether this limit holds in more exotic circumstances, such as exoplanet silicate atmospheres with temperatures in the thousands of kelvins \cite[e.g.,][]{kang2021escaping}, to future work.}, the factor $L/(RT)$ ranges from about 25--50, while the factor $L^2/(c_p R_v T^2)$ ranges from about 100--500, in both cases dominating over $a$, which typically takes on values of $\mathcal{O}(0.1)-\mathcal{O}(1)$ \citep{romps2016cape}. Hence the entraining lapse rate asymptotes to $\Gamma \simeq g T/L$ --- which, as in the dry limit, is independent of the entrainment rate, and also equal to the moist limit of $\Gamma_m$. To summarize, equation (\ref{eq:gamma_ent}) suggests that we should expect minimal CAPE in both the dry and moist limits, with a peak in between for the intermediate values of $q_v^*$ that allow entrainment to drive a wedge between $\Gamma$ and $\Gamma_m$. 

\begin{figure}[ht!]
\centerline{\includegraphics[width=\textwidth]{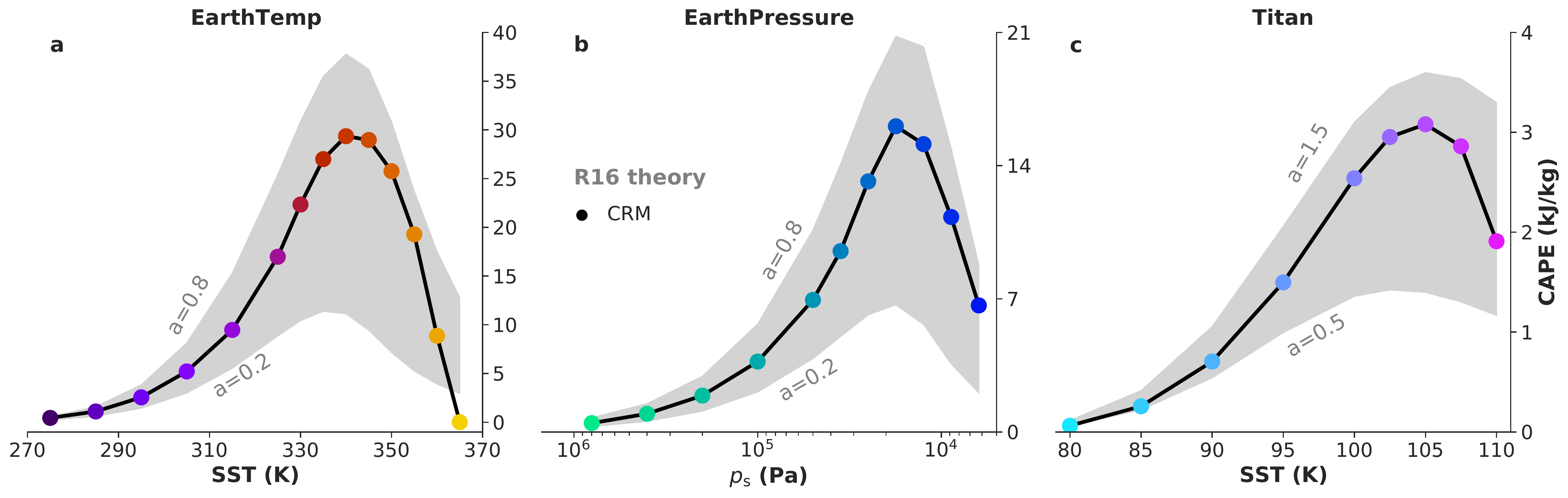}}
\caption{CAPE from the (a) \emph{EarthTemp}, (b) \emph{EarthPressure}, and (c) \emph{Titan} experiments from the convection-resolving simulations (colored markers) and from the analytical predictions for CAPE given by equation (\ref{eq:CAPE_R16}) (gray shaded areas). The analytical predictions are shown for values of the bulk-plume parameter $a$ ranging from 0.2 to 0.8 for the Earth-like experiments, and for $a$ ranging from 0.5 to 1.5 for the Titan-like experiment. The analytical solutions are initialized with the temperature and humidity of the lifted condensation level from the corresponding simulation.}
\label{fig:R16}
\end{figure}

Quantitatively, the magnitude and atmospheric humidity of the peak in CAPE can be predicted using the analytic formalism of \cite{romps2016cape}. Figure \ref{fig:R16} shows analytical predictions for CAPE using the theory of R16 (reproduced in our equation \ref{eq:CAPE_R16}), in comparison to the results from our convection-resolving simulations. The theory clearly captures the boom-bust evolution of CAPE in all three core experiments, providing the theoretical bridge between the ``warming-driven invigoration'' and the ``gentle pure-steam limit'' regimes. While R16 applied their analytic theory for CAPE to the context of surface warming on Earth, here we have shown that the same essential physics explains convective vigor in convection-resolving simulations with varied surface temperature, varied surface pressure, and with Titan-like thermodynamics. The notion that entraining and adiabatic temperature profiles converge in non-dilute atmospheres resembles the arguments given by \cite{Ding2016} and \cite{Pierrehumbert2016}, although our results suggest that this physics begins to operate well before the steam limit is reached. \cite{Ding2016} and \cite{Pierrehumbert2016} also argued that environmental relative humidity should approach 1 as the atmosphere becomes increasingly non-dilute, a prediction that is confirmed by our results (Fig. \ref{fig:RH}).

In addition to explaining the behavior of CAPE in our convection-resolving simulations, the theory of R16 provides clarity on which planetary parameters control CAPE. An interesting attribute of equation \ref{eq:CAPE_R16} is that it has no explicit dependence on a planet's gravitational constant, $g$. Physically, this can be understood as a cancellation between two factors: 1) for a given parcel temperature anomaly, there is a direct proportionality between the parcel's buoyancy and $g$; and 2) for a given surface temperature and tropopause temperature, there is an \emph{inverse} proportionality between the geometric depth of the atmosphere and $g$; hence reducing gravity lowers the integrand in equation \ref{eq:CAPE} but increases the domain of vertical integration by a compensating amount.

To test this prediction of an insensitivity to $g$, we ran a subset of surface temperature cases from the \emph{EarthTemp} experiment with either enhanced ($g=25$ m/s$^2$) or reduced ($g=3.5$ m/s$^2$) gravitational constant (the \emph{VaryGrav} experiment); these values of $g$ approximately cover surface conditions for solar system planets ranging in mass from Mercury to Jupiter. We find a weak dependence of CAPE on $g$ (i.e., CAPE varies by a factor of $\simeq$2 when $g$ varies by a factor of $\simeq$7), in rough accordance with the theoretical prediction, and also find a correspondingly small sensitivity in our metrics of actual convective vigor to $g$. The small variations in CAPE and convective vigor likely result from the effect of $g$ on the nature of turbulence in the simulations, which would affect both the entrainment rate that enters into the theory for CAPE (through the bulk-plume parameter $a$) and the drag experienced by actual convecting parcels. 

These experiments with varied gravity also afford an opportunity to test the sensitivity of our results to the precipitation fall speed parameter, which some studies have suggested is a key control on updraft speeds \citep{Parodi2009}. Assuming a conservative\footnote{Theoretical results for monodisperse droplets predict a square-root dependence of raindrop terminal velocity on $g$ \citep{Loftus2021}.} linear dependence of precipitation fall speed on $g$, we re-ran the cases with $g=3.5$ m/s$^2$ and $g=25$ m/s$^2$ with the precipitation fall speeds modified to 2.8 m/s and 20 m/s, respectively. We found this had a minimal effect, with changes in CAPE and convective vigor generally limited to $\pm5$\%.

\section{Discussion}\label{sec:disc}

Using an idealized convection-resolving model, we have demonstrated that convective vigor is expected to peak at intermediate concentrations of the condensing substance in a diversity of planetary circumstances. However, more work is needed to build our results into a universal understanding of convective vigor. Our simulations of local radiative-convective equilibrium are highly idealized, and many questions remain about how the peak in convective vigor with respect to atmospheric humidity would manifest in more realistic modeling setups that include coupling to large-scale circulations \citep{Pierrehumbert2016} or a diurnal cycle. Another promising avenue for extension concerns our Titan-like simulations, for which we assumed a limitless supply of surface evaporation. In reality, the surface of Titan is quite arid \citep{Schneider2012}, and future work could explore convective vigor in the regime where the atmospheric condensible inventory is comparable to the total inventory.

Additionally, although we have successfully applied the theory of R16 to our results, that theory is limited in its general applicability because it approximates the specific gas constant and heat capacity of moist air by those of the dry component. This is a tolerable source of error when the dry and condensing components do not differ too much in molar mass, as in the case of H$_2$O condensing in an N$_2$/O$_2$ mixture or CH$_4$ condensing in N$_2$, but this approximation breaks down when the background gas is very light (e.g., mainly H$_2$). In that case, the atmospheric scale height can collapse with warming as the atmosphere comes to be dominated by the relatively heavier condensing component \citep{Koll2019}. Since many planets form with a primordial hydrogen envelope, this is an important class of atmospheres to which the R16 theory and our simulation results may not apply. Additional theoretical and computational work could investigate this regime, for which the ``virtual'' effects of compositional differences on buoyancy may be crucial.

In additional to building fundamental understanding of moist convection, our results may also have implications for planetary evolution on long time scales. The rate at which a terrestrial planet loses water depends on stratospheric humidity, because water transported to the stratosphere becomes vulnerable to photolysis and subsequent loss of H to space \citep{Kasting1988}. Purely thermodynamic arguments suggest that stratospheric moistening in planetary atmospheres depends on both surface temperature and surface pressure, with the transition to a moist stratosphere occurring at intermediate humidities \citep{Wordsworth2013}. However, on present-day Earth, injection of water into the stratosphere by intense convective storms plays an important role in setting the average stratospheric water content \citep{corti2008unprecedented}. It may be that in real atmospheres, the increase in convective vigor at intermediate humidity causes the ``moist greenhouse" transition \citep{Kasting1988} to be approached more rapidly than either one-dimensional radiative-convective models or 3D general circulation model simulations \citep[e.g., ][]{leconte2013increased} would suggest. Further research using models that couple convection to large-scale dynamical and radiative processes is required to investigate this possibility.

Furthermore, because the rate of lightning strikes in planetary atmospheres is believed to depend in part on convective vigor, these results may also have important implications for lightning-driven atmospheric chemistry. \cite{Romps2014} proposed that the lightning flash rate on modern Earth is proportional to the product of CAPE and precipitation rate. If this relation is robust across wide ranges of planetary conditions, it implies that the importance of lightning chemistry would be strongly enhanced in atmospheres with intermediate specific humidity. Because both surface temperature and atmospheric pressure may have varied significantly on Earth in the Hadean, this has interesting implications for the rate of lightning-driven formation of important prebiotic molecules such as HCN during this period \citep{chameides1981rates,ardaseva2017lightning}.

\begin{acknowledgments}
\textbf{Data availability:} Cloud-resolving model output and the code that generates the figures in this manuscript is available at \url{https://doi.org/10.5281/zenodo.7331932}.
\end{acknowledgments}

%

\vspace{5mm}





\appendix

\section{Analytical expression for CAPE from Romps (2016)}
The solutions of R16 yield a closed-form expression for the CAPE of an atmosphere in RCE:

\begin{multline} \label{eq:CAPE_R16}
    \mathrm{CAPE} = \frac{R}{2f}\{W(y_a)[2 - 2f(T_s - T_t) + W(y_a)] - W(e^{-f(T_s-T_t)}y_a)[2 + W(e^{-f(T_s - T_t)}y_a)] \\ - W(y_0)2 - 2f(T_s - T_t) + W(y_0)] + W(e^{-f(T_s-T_t)}y_0)[2 + W(e^{-f(T_s - T_t)}y_0)]\}.
\end{multline}

\noindent where $W$ is the Lambert W function defined by $W(xe^x)=x$, and where

\begin{align}
    f   &= \frac{L}{R_v T_0^2} - \frac{c_p}{R T_0}, \quad \quad \quad \mathrm{and} \label{eq:f}\\
    y_a &= \frac{L q_{vs}^*}{(1+a)R T_0}\exp\left[\frac{L q_{vs}^*}{(1+a)R T_0}\right] \label{eq:y_a}.
\end{align}

\noindent Here, $R$ (J/kg/K) is the specific gas constant of dry (background) air, $R_v$ (J/kg/K) is the specific gas constant of the condensible vapor, $L$ (J/kg) is the latent heat of condensation (assumed constant), $c_p$ (J/kg/K) is the specific heat capacity at constant pressure of dry air, $q_{vs}^*$ (kg/kg) is the saturation specific humidity at the surface, and $T_0$ (K) is a constant reference temperature chosen to be the average of the surface temperature $T_s$ and the tropopause temperature $T_t$. The dimensionless parameter $a$ characterizes the bulk-plume convection. In equation \ref{eq:CAPE_R16}, $y_0$ is given by equation \ref{eq:y_a} with $a=0$.

The expression for CAPE given above depends only on known physical constants, observable planetary conditions, and a summary parameter characterizing the bulk-plume convection. The bulk-plume parameter $a$ is proportional to the entrainment rate\footnote{Specifically, $a=\epsilon\mathrm{PE}/\gamma$, where $\epsilon$ (1/m) is the fractional entrainment rate, the precipitation efficiency PE is defined as the ratio of net condensation to gross condensation (assumed constant throughout the troposphere), and $\gamma \equiv - \partial_z \log{q_v^*}$ (1/m) is the saturation water-vapor lapse rate. One of the assumptions made by R16 for analytical tractability is that $\epsilon\propto\gamma$, so that $a$ is constant.}, and it is easy to verify that setting $a=0$ (the limit of non-entraining convection) produces a moist-adiabatic atmosphere with zero CAPE, as expected. We can also verify that equation \ref{eq:CAPE_R16} makes quantitatively accurate predictions for Earth's tropics today. We set $a=0.2$, which corresponds to typical values of precipitation efficiency and entrainment rate as diagnosed in cloud-resolving simulations, as discussed in  R16. Additionally using $T_s=300$ K, $q_{vs}^* = 20$ g/kg, $L=2.5\times10^6$ J/kg, and $T_t=200$ K, equation \ref{eq:CAPE_R16} predicts $\mathrm{CAPE}\simeq2500$ J/kg, which is indeed a typical value for convectively-active parts of Earth's tropics \citep{Riemann-Campe2009}.

\restartappendixnumbering
\section{Supplemental figures}

\begin{figure}[h!]
 \centerline{\includegraphics[width=4in]{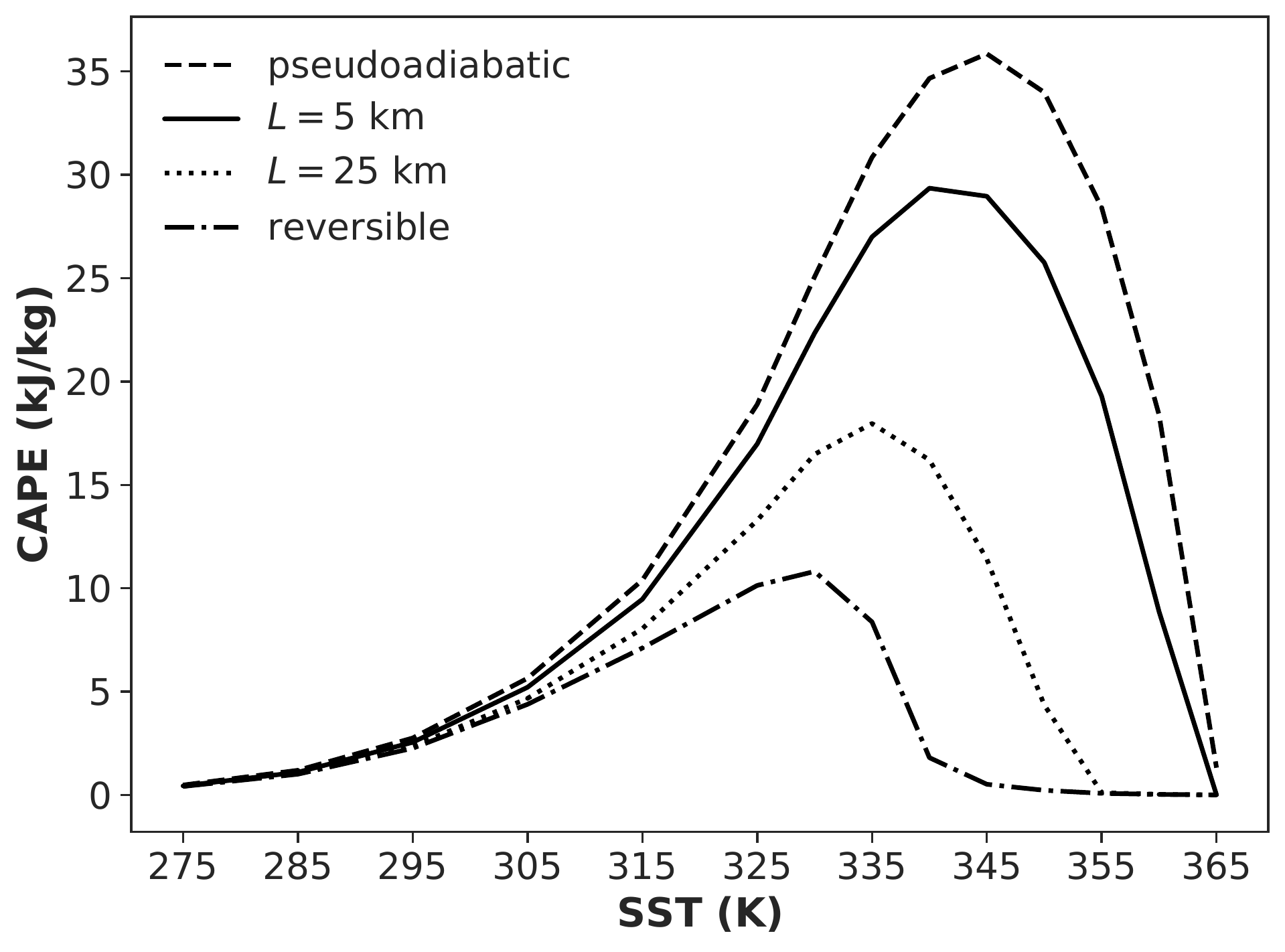}}
\caption{The effect of varying condensate fallout assumptions on parcel lifting calculations for computation of CAPE. The condensed water mass fraction is assumed to have a sink term due to fallout that manifests as an exponential decay with height, on a length scale $L$, during each discrete lifting step. The main text figures use $L=5$ km, but here we also show the cases $L\rightarrow0$ (pseudoadiabatic), $L=25$ km, and $L\rightarrow\infty$ (reversible). In all cases there is a peak in CAPE.}\label{fig:fallout}
\end{figure}

\begin{figure}[h!]
 \centerline{\includegraphics[width=6in]{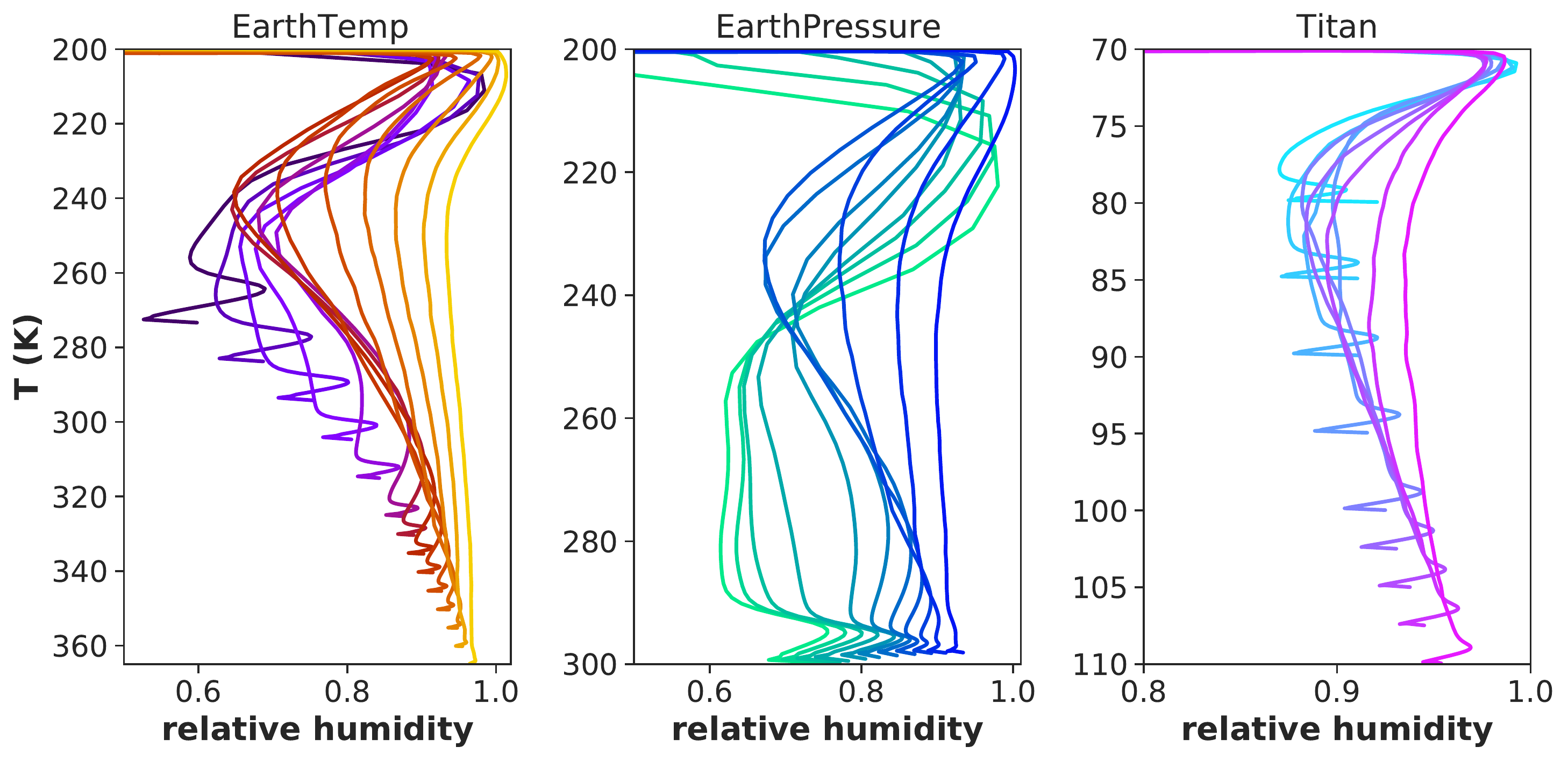}}
\caption{Profiles of environmental relative humidity in the (left) \emph{EarthTemp}, (center) \emph{EarthPressure}, and (right) \emph{Titan} experiments. Relative humidity is plotted as a function of mean temperature in the troposphere, as in \cite{Romps2014a}. The line colors are as in the main text figures, with relative humidity generally increasing as the condensible species becomes more prevalent.}\label{fig:RH}
\end{figure}


\bibliography{nondilute}{}
\bibliographystyle{aasjournal}



\end{document}